\newif\ifdraft
\newif\iffull
\newif\ifcomment
\newif\iflatexdiff
\newif\ifbibtex
\newif\ifpreprint
\preprinttrue    
\fulltrue        
\def\dvers{v0.36}

\ifpreprint

\fi
\def\stitle{Charge correlations using the balance function in Pb--Pb collisions}
\ifpreprint
\documentclass[ALICE,manyauthors,12pt]{cernphprep}
\usepackage[comma,square,numbers,sort&compress]{natbib}
\usepackage{hyperref}
\else
\documentclass[final,3p,12pt]{elsarticle}
\biboptions{comma,square,numbers,sort&compress}
\usepackage{hyperref}
\fi
\usepackage{graphicx}  
\usepackage{dcolumn}   
\usepackage{bm}        
\usepackage{amssymb}   
\usepackage{amsfonts}
\usepackage{graphics}
\usepackage{grffile}   
\usepackage{epsfig}
\usepackage{units}
\usepackage[usenames]{color}
\usepackage[normalem]{ulem} 
\usepackage{color}
\usepackage[utf8]{inputenc}
\usepackage[T1]{fontenc}
\iflatexdiff
\RequirePackage{color}\definecolor{RED}{rgb}{1,0,0}\definecolor{BLUE}{rgb}{0,0,1}

\fi
\ifdraft
\usepackage{lineno}
\linenumbers
\modulolinenumbers[5]
\usepackage{fancyhdr}
\else
\fi

\newcommand{\pt}{$p_{\rm{T}}$}
\newcounter{vers}
\newcommand{\ITS}{\rm{ITS}}
\newcommand{\SPD}{\rm{SPD}}
\newcommand{\SDD}{\rm{SDD}}
\newcommand{\SSD}{\rm{SSD}}
\newcommand{\TPC}{\rm{TPC}}

\newcommand{\VZERO}{\rm{VZERO}}

\newcommand{\deta}{$\Delta\eta$}
\newcommand{\dphi}{$\Delta\varphi$}
\newcommand{\wdeta}{$\langle \Delta\eta \rangle$}
\newcommand{\wdphi}{$\langle \Delta\varphi \rangle$}
\begin{document}
\newlength{\figlen}
\setlength{\figlen}{\linewidth}
\ifpreprint
\setlength{\figlen}{0.95\linewidth}
\begin{titlepage}
\PHnumber{2012-371}                   
\PHdate{Dec 20, 2012}                  
\title{Charge correlations using the balance function in Pb--Pb collisions at $\mathbf{\sqrt{s_{\rm NN}} = 2.76}$~TeV}
\ShortTitle{\stitle}
\Collaboration{ALICE Collaboration%
         \thanks{See Appendix~\ref{app:collab} for the list of collaboration members}}
\ShortAuthor{ALICE Collaboration} 
\ifdraft
\begin{center}
\today\\ \color{red}DRAFT \dvers\ \hspace{0.3cm} \$Revision: 148 $\color{white}:$\$\color{black}\vspace{0.3cm}
\end{center}
\fi
\else
\begin{frontmatter}
\title{\dtitle}
\iffull



\begingroup
\small
\begin{flushleft}
B.~Abelev\Irefn{org1234}\And
J.~Adam\Irefn{org1274}\And
D.~Adamov\'{a}\Irefn{org1283}\And
A.M.~Adare\Irefn{org1260}\And
M.M.~Aggarwal\Irefn{org1157}\And
G.~Aglieri~Rinella\Irefn{org1192}\And
M.~Agnello\Irefn{org1313}\textsuperscript{,}\Irefn{org1017688}\And
A.G.~Agocs\Irefn{org1143}\And
A.~Agostinelli\Irefn{org1132}\And
Z.~Ahammed\Irefn{org1225}
% ...
\renewcommand\labelenumi{\textsuperscript{\theenumi}~}

\section*{Affiliation notes}
\renewcommand\theenumi{\roman{enumi}}
\begin{Authlist}
\item \Adef{0}Deceased
\item \Adef{idp4109216}{Also at: M.V.Lomonosov Moscow State University, D.V.Skobeltsyn Institute of Nuclear Physics, Moscow, Russia}
\item \Adef{idp4393328}{Also at: University of Belgrade, Faculty of Physics and "Vin\\v{c}a" Institute of Nuclear Sciences, Belgrade, Serbia}
\item \Adef{idp5445312}{Also at: Institute of Theoretical Physics, University of Wroclaw, Wroclaw, Poland}
\end{Authlist}

\section*{Collaboration Institutes}
\renewcommand\theenumi{\arabic{enumi}~}
\begin{Authlist}

\item \Idef{org1332}A. I. Alikhanyan National Science Laboratory (Yerevan Physics Institute) Foundation, Yerevan, Armenia
\item \Idef{org1279}Benem\'{e}rita Universidad Aut\'{o}noma de Puebla, Puebla, Mexico
\item \Idef{org1220}Bogolyubov Institute for Theoretical Physics, Kiev, Ukraine
\item \Idef{org20959}Bose Institute, Department of Physics and Centre for Astroparticle Physics and Space Science (CAPSS), Kolkata, India
\item \Idef{org1262}Budker Institute for Nuclear Physics, Novosibirsk, Russia
\item \Idef{org1292}California Polytechnic State University, San Luis Obispo, California, United States
\item \Idef{org1329}Central China Normal University, Wuhan, China
\item \Idef{org14939}Centre de Calcul de l'IN2P3, Villeurbanne, France
\item \Idef{org1197}Centro de Aplicaciones Tecnol\'{o}gicas y Desarrollo Nuclear (CEADEN), Havana, Cuba
\item \Idef{org1242}Centro de Investigaciones Energ\'{e}ticas Medioambientales y Tecnol\'{o}gicas (CIEMAT), Madrid, Spain
\item \Idef{org1244}Centro de Investigaci\'{o}n y de Estudios Avanzados (CINVESTAV), Mexico City and M\'{e}rida, Mexico
\item \Idef{org1335}Centro Fermi - Museo Storico della Fisica e Centro Studi e Ricerche ``Enrico Fermi'', Rome, Italy
\item \Idef{org17347}Chicago State University, Chicago, United States
\item \Idef{org1288}Commissariat \`{a} l'Energie Atomique, IRFU, Saclay, France
\item \Idef{org15782}COMSATS Institute of Information Technology (CIIT), Islamabad, Pakistan
\item \Idef{org1294}Departamento de F\'{\i}sica de Part\'{\i}culas and IGFAE, Universidad de Santiago de Compostela, Santiago de Compostela, Spain
\item \Idef{org1106}Department of Physics Aligarh Muslim University, Aligarh, India
\item \Idef{org1121}Department of Physics and Technology, University of Bergen, Bergen, Norway
\item \Idef{org1162}Department of Physics, Ohio State University, Columbus, Ohio, United States
\item \Idef{org1300}Department of Physics, Sejong University, Seoul, South Korea
\item \Idef{org1268}Department of Physics, University of Oslo, Oslo, Norway
\item \Idef{org1312}Dipartimento di Fisica dell'Universit\`{a} and Sezione INFN, Turin, Italy
\item \Idef{org1145}Dipartimento di Fisica dell'Universit\`{a} and Sezione INFN, Cagliari, Italy
\item \Idef{org1315}Dipartimento di Fisica dell'Universit\`{a} and Sezione INFN, Trieste, Italy
\item \Idef{org1285}Dipartimento di Fisica dell'Universit\`{a} `La Sapienza' and Sezione INFN, Rome, Italy
\item \Idef{org1154}Dipartimento di Fisica e Astronomia dell'Universit\`{a} and Sezione INFN, Catania, Italy
\item \Idef{org1132}Dipartimento di Fisica e Astronomia dell'Universit\`{a} and Sezione INFN, Bologna, Italy
\item \Idef{org1270}Dipartimento di Fisica e Astronomia dell'Universit\`{a} and Sezione INFN, Padova, Italy
\item \Idef{org1290}Dipartimento di Fisica `E.R.~Caianiello' dell'Universit\`{a} and Gruppo Collegato INFN, Salerno, Italy
\item \Idef{org1103}Dipartimento di Scienze e Innovazione Tecnologica dell'Universit\`{a} del Piemonte Orientale and Gruppo Collegato INFN, Alessandria, Italy
\item \Idef{org1114}Dipartimento Interateneo di Fisica `M.~Merlin' and Sezione INFN, Bari, Italy
\item \Idef{org1237}Division of Experimental High Energy Physics, University of Lund, Lund, Sweden
\item \Idef{org1192}European Organization for Nuclear Research (CERN), Geneva, Switzerland
\item \Idef{org1227}Fachhochschule K\"{o}ln, K\"{o}ln, Germany
\item \Idef{org1122}Faculty of Engineering, Bergen University College, Bergen, Norway
\item \Idef{org1136}Faculty of Mathematics, Physics and Informatics, Comenius University, Bratislava, Slovakia
\item \Idef{org1274}Faculty of Nuclear Sciences and Physical Engineering, Czech Technical University in Prague, Prague, Czech Republic
\item \Idef{org1229}Faculty of Science, P.J.~\v{S}af\'{a}rik University, Ko\v{s}ice, Slovakia
\item \Idef{org1184}Frankfurt Institute for Advanced Studies, Johann Wolfgang Goethe-Universit\"{a}t Frankfurt, Frankfurt, Germany
\item \Idef{org1215}Gangneung-Wonju National University, Gangneung, South Korea
\item \Idef{org20958}Gauhati University, Department of Physics, Guwahati, India
\item \Idef{org1212}Helsinki Institute of Physics (HIP) and University of Jyv\"{a}skyl\"{a}, Jyv\"{a}skyl\"{a}, Finland
\item \Idef{org1203}Hiroshima University, Hiroshima, Japan
\item \Idef{org1254}Indian Institute of Technology Bombay (IIT), Mumbai, India
\item \Idef{org36378}Indian Institute of Technology Indore, Indore, India (IITI)
\item \Idef{org1266}Institut de Physique Nucl\'{e}aire d'Orsay (IPNO), Universit\'{e} Paris-Sud, CNRS-IN2P3, Orsay, France
\item \Idef{org1277}Institute for High Energy Physics, Protvino, Russia
\item \Idef{org1249}Institute for Nuclear Research, Academy of Sciences, Moscow, Russia
\item \Idef{org1320}Nikhef, National Institute for Subatomic Physics and Institute for Subatomic Physics of Utrecht University, Utrecht, Netherlands
\item \Idef{org1250}Institute for Theoretical and Experimental Physics, Moscow, Russia
\item \Idef{org1230}Institute of Experimental Physics, Slovak Academy of Sciences, Ko\v{s}ice, Slovakia
\item \Idef{org1127}Institute of Physics, Bhubaneswar, India
\item \Idef{org1275}Institute of Physics, Academy of Sciences of the Czech Republic, Prague, Czech Republic
\item \Idef{org1139}Institute of Space Sciences (ISS), Bucharest, Romania
\item \Idef{org27399}Institut f\"{u}r Informatik, Johann Wolfgang Goethe-Universit\"{a}t Frankfurt, Frankfurt, Germany
\item \Idef{org1185}Institut f\"{u}r Kernphysik, Johann Wolfgang Goethe-Universit\"{a}t Frankfurt, Frankfurt, Germany
\item \Idef{org1177}Institut f\"{u}r Kernphysik, Technische Universit\"{a}t Darmstadt, Darmstadt, Germany
\item \Idef{org1256}Institut f\"{u}r Kernphysik, Westf\"{a}lische Wilhelms-Universit\"{a}t M\"{u}nster, M\"{u}nster, Germany
\item \Idef{org1246}Instituto de Ciencias Nucleares, Universidad Nacional Aut\'{o}noma de M\'{e}xico, Mexico City, Mexico
\item \Idef{org1247}Instituto de F\'{\i}sica, Universidad Nacional Aut\'{o}noma de M\'{e}xico, Mexico City, Mexico
\item \Idef{org1308}Institut Pluridisciplinaire Hubert Curien (IPHC), Universit\'{e} de Strasbourg, CNRS-IN2P3, Strasbourg, France
\item \Idef{org1182}Joint Institute for Nuclear Research (JINR), Dubna, Russia
\item \Idef{org1199}Kirchhoff-Institut f\"{u}r Physik, Ruprecht-Karls-Universit\"{a}t Heidelberg, Heidelberg, Germany
\item \Idef{org20954}Korea Institute of Science and Technology Information, Daejeon, South Korea
\item \Idef{org1017642}KTO Karatay University, Konya, Turkey
\item \Idef{org1160}Laboratoire de Physique Corpusculaire (LPC), Clermont Universit\'{e}, Universit\'{e} Blaise Pascal, CNRS--IN2P3, Clermont-Ferrand, France
\item \Idef{org1194}Laboratoire de Physique Subatomique et de Cosmologie (LPSC), Universit\'{e} Joseph Fourier, CNRS-IN2P3, Institut Polytechnique de Grenoble, Grenoble, France
\item \Idef{org1187}Laboratori Nazionali di Frascati, INFN, Frascati, Italy
\item \Idef{org1232}Laboratori Nazionali di Legnaro, INFN, Legnaro, Italy
\item \Idef{org1125}Lawrence Berkeley National Laboratory, Berkeley, California, United States
\item \Idef{org1234}Lawrence Livermore National Laboratory, Livermore, California, United States
\item \Idef{org1251}Moscow Engineering Physics Institute, Moscow, Russia
\item \Idef{org1322}National Centre for Nuclear Studies, Warsaw, Poland
\item \Idef{org1140}National Institute for Physics and Nuclear Engineering, Bucharest, Romania
\item \Idef{org1017626}National Institute of Science Education and Research, Bhubaneswar, India
\item \Idef{org1165}Niels Bohr Institute, University of Copenhagen, Copenhagen, Denmark
\item \Idef{org1109}Nikhef, National Institute for Subatomic Physics, Amsterdam, Netherlands
\item \Idef{org1283}Nuclear Physics Institute, Academy of Sciences of the Czech Republic, \v{R}e\v{z} u Prahy, Czech Republic
\item \Idef{org1264}Oak Ridge National Laboratory, Oak Ridge, Tennessee, United States
\item \Idef{org1189}Petersburg Nuclear Physics Institute, Gatchina, Russia
\item \Idef{org1170}Physics Department, Creighton University, Omaha, Nebraska, United States
\item \Idef{org1157}Physics Department, Panjab University, Chandigarh, India
\item \Idef{org1112}Physics Department, University of Athens, Athens, Greece
\item \Idef{org1152}Physics Department, University of Cape Town and  iThemba LABS, National Research Foundation, Somerset West, South Africa
\item \Idef{org1209}Physics Department, University of Jammu, Jammu, India
\item \Idef{org1207}Physics Department, University of Rajasthan, Jaipur, India
\item \Idef{org1200}Physikalisches Institut, Ruprecht-Karls-Universit\"{a}t Heidelberg, Heidelberg, Germany
\item \Idef{org1017688}Politecnico di Torino, Turin, Italy
\item \Idef{org1325}Purdue University, West Lafayette, Indiana, United States
\item \Idef{org1281}Pusan National University, Pusan, South Korea
\item \Idef{org1176}Research Division and ExtreMe Matter Institute EMMI, GSI Helmholtzzentrum f\"ur Schwerionenforschung, Darmstadt, Germany
\item \Idef{org1334}Rudjer Bo\v{s}kovi\'{c} Institute, Zagreb, Croatia
\item \Idef{org1298}Russian Federal Nuclear Center (VNIIEF), Sarov, Russia
\item \Idef{org1252}Russian Research Centre Kurchatov Institute, Moscow, Russia
\item \Idef{org1224}Saha Institute of Nuclear Physics, Kolkata, India
\item \Idef{org1130}School of Physics and Astronomy, University of Birmingham, Birmingham, United Kingdom
\item \Idef{org1338}Secci\'{o}n F\'{\i}sica, Departamento de Ciencias, Pontificia Universidad Cat\'{o}lica del Per\'{u}, Lima, Peru
\item \Idef{org1133}Sezione INFN, Bologna, Italy
\item \Idef{org1271}Sezione INFN, Padova, Italy
\item \Idef{org1286}Sezione INFN, Rome, Italy
\item \Idef{org1146}Sezione INFN, Cagliari, Italy
\item \Idef{org1313}Sezione INFN, Turin, Italy
\item \Idef{org1316}Sezione INFN, Trieste, Italy
\item \Idef{org1115}Sezione INFN, Bari, Italy
\item \Idef{org1155}Sezione INFN, Catania, Italy
\item \Idef{org36377}Nuclear Physics Group, STFC Daresbury Laboratory, Daresbury, United Kingdom
\item \Idef{org1258}SUBATECH, Ecole des Mines de Nantes, Universit\'{e} de Nantes, CNRS-IN2P3, Nantes, France
\item \Idef{org35706}Suranaree University of Technology, Nakhon Ratchasima, Thailand
\item \Idef{org1304}Technical University of Split FESB, Split, Croatia
\item \Idef{org1168}The Henryk Niewodniczanski Institute of Nuclear Physics, Polish Academy of Sciences, Cracow, Poland
\item \Idef{org17361}The University of Texas at Austin, Physics Department, Austin, TX, United States
\item \Idef{org1173}Universidad Aut\'{o}noma de Sinaloa, Culiac\'{a}n, Mexico
\item \Idef{org1296}Universidade de S\~{a}o Paulo (USP), S\~{a}o Paulo, Brazil
\item \Idef{org1149}Universidade Estadual de Campinas (UNICAMP), Campinas, Brazil
\item \Idef{org1239}Universit\'{e} de Lyon, Universit\'{e} Lyon 1, CNRS/IN2P3, IPN-Lyon, Villeurbanne, France
\item \Idef{org1205}University of Houston, Houston, Texas, United States
\item \Idef{org20371}University of Technology and Austrian Academy of Sciences, Vienna, Austria
\item \Idef{org1222}University of Tennessee, Knoxville, Tennessee, United States
\item \Idef{org1310}University of Tokyo, Tokyo, Japan
\item \Idef{org1318}University of Tsukuba, Tsukuba, Japan
\item \Idef{org21360}Eberhard Karls Universit\"{a}t T\"{u}bingen, T\"{u}bingen, Germany
\item \Idef{org1225}Variable Energy Cyclotron Centre, Kolkata, India
\item \Idef{org1306}V.~Fock Institute for Physics, St. Petersburg State University, St. Petersburg, Russia
\item \Idef{org1323}Warsaw University of Technology, Warsaw, Poland
\item \Idef{org1179}Wayne State University, Detroit, Michigan, United States
\item \Idef{org1143}Wigner Research Centre for Physics, Hungarian Academy of Sciences, Budapest, Hungary
\item \Idef{org1260}Yale University, New Haven, Connecticut, United States
\item \Idef{org15649}Yildiz Technical University, Istanbul, Turkey
\item \Idef{org1301}Yonsei University, Seoul, South Korea
\item \Idef{org1327}Zentrum f\"{u}r Technologietransfer und Telekommunikation (ZTT), Fachhochschule Worms, Worms, Germany
\end{Authlist}
\endgroup

\else
\ifdraft
\author{ALICE Collaboration \\ \vspace{0.3cm}
\today\\ \color{red}DRAFT \dvers\ \hspace{0.3cm} \$Revision: 148 $\color{white}:$\$\color{black}}
\else
\author{ALICE Collaboration}
\fi
\fi
\fi
\begin{abstract}
In high-energy heavy-ion collisions, the correlations between the emitted particles can be used as a probe to gain 
insight into the charge creation mechanisms. In this Letter, we report the first results of such studies using the electric 
charge balance function in the relative pseudorapidity (\deta) and azimuthal angle (\dphi) in Pb--Pb collisions at 
$\sqrt{s_{\rm NN}}=2.76$~TeV with the ALICE detector at the Large Hadron Collider.  The width of the balance function 
decreases with growing centrality (i.e. for more central collisions) in both projections. 
This centrality dependence is not reproduced by HIJING, while AMPT, a model which incorporates strings and parton rescattering, 
exhibits qualitative agreement with the measured correlations in $\Delta \varphi$ but fails to describe the correlations in \deta. 
A thermal blast-wave model incorporating local charge conservation and tuned to describe the \pt~spectra and v$_2$ measurements 
reported by ALICE, is used to fit the centrality dependence of the width of the balance function and to extract the average separation 
of balancing charges at freeze-out. 
The comparison of our results with measurements at lower energies reveals an ordering with $\sqrt{s_{\rm NN}}$: 
the balance functions become narrower with increasing energy for all centralities. 
This is consistent with the effect of larger radial flow at the LHC energies but also with the late stage creation scenario of balancing charges. 
However, the relative decrease of the balance function widths in \deta~and \dphi~with centrality from the highest 
SPS to the LHC energy exhibits only small differences. This observation cannot be interpreted solely within the  
framework where the majority of the charge is produced at a later stage in the evolution of the heavy-ion collision.
\ifdraft
\ifpreprint
\end{abstract}
\end{titlepage}
\else
\end{abstract}
\end{frontmatter}
\newpage
\fi
\fi
\ifdraft
\thispagestyle{fancyplain}
\else
\end{abstract}
\ifpreprint
\end{titlepage}
\else
\end{frontmatter}
\fi
\fi
\setcounter{page}{2}

\section{Introduction}
\label{Section:Intro}
According to Quantum ChromoDynamics (QCD), the theory that describes the strong interaction, 
at sufficiently high energy densities and temperatures, a new phase of matter exists in which the constituents, the quarks and the gluons, are deconfined \cite{Ref:QCD}. This new state of matter is called the Quark Gluon Plasma (QGP). Its creation in 
the laboratory, the corresponding verification of its existence and the subsequent study of its 
properties are the main goals of the ultrarelativistic heavy-ion collision programs. Convincing experimental 
evidences for the existence of a deconfined phase have been published already at RHIC energies 
\cite{Ref:RHICQGP}. Recently, the first experimental results from the heavy-ion program of the LHC experiments provided additional indication \cite{Ref:AliceFlow,Ref:LHCHighPt} for the existence of this state of matter at this new energy regime.

Among the different observables, such as the anisotropic flow \cite{Ref:AliceFlow} or the energy 
loss of high transverse momentum particles \cite{Ref:LHCHighPt}, the charge balance 
functions are suggested to be sensitive probes of the properties of the system, providing valuable insight into the charge creation mechanism and can be 
used to address fundamental questions concerning hadronization in heavy-ion 
collisions \cite{Ref:BalanceFunctionTheory}.

The system that is produced in a heavy-ion collision undergoes an expansion, during which it exhibits collective behavior and can be 
described in terms of hydrodynamics \cite{Ref:HeinzHydro}. A pair of particles of opposite charge 
that is created during this stage is subject to the collective motion of the system, which transforms the correlations in coordinate space into correlations in momentum space. The subsequent rescattering phase after the hadronization 
will also affect the final measured degree of correlation. The balance function being a sensitive probe of the balancing charge distribution in momentum space, quantifies these effects. The final degree of 
correlation is reflected in the balance function distribution and consequently in its width. It was suggested in \cite{Ref:BalanceFunctionTheory} that narrow 
distributions correspond to a system that consists of particles that are created close to the end of the evolution. It was also suggested that a larger width may signal the creation of balancing charges at the first 
stages of the system's evolution \cite{Ref:BalanceFunctionTheory}.

The balance function reflects the strength of correlation between a particle in a bin $P_1$ in momentum space and the accompanying (balancing) particle of opposite charge with momentum $P_2$. The general 
definition is given in Eq.~\ref{Eq:BFGeneralDefinition}:
\begin{equation}
B_{ab}(P_2,P_1) = \frac{1}{2}\Big(C_{ab}(P_2,P_1) + C_{ba}(P_2,P_1)
- C_{bb}(P_2,P_1) - C_{aa}(P_2,P_1)\Big), \label{Eq:BFGeneralDefinition}
\end{equation}

\noindent where $C_{ab}(P_2,P_1) = N_{ab}(P_2,P_1)/N_b(P_1)$ is the distribution of pairs of particles, of type $a$ and $b$, with momenta $P_2$ and $P_1$, respectively, normalized 
to the number of particles $b$. Particles $a$ and $b$ could come from different particle species (e.g. $\pi^{+}$--$\pi^{-}$, 
K$^{+}$--K$^{-}$, p--$\overline{\rm p}$). In this Letter, $a$ refers to all positive and $b$ to all negative particles. 
This analysis is performed for both particles in the pseudorapidity intervals  $|\eta| < 0.8$. We assume that the balance function is invariant over pseudorapidity in this region, and  report the results in terms of the relative pseudorapidity  $\Delta \eta = \eta_b - \eta_a$ and the relative azimuthal angle $\Delta \varphi = \varphi_b - \varphi_a$, by averaging the balance function over the position of one of the particles (similar equation is used for $B(\Delta \varphi)$): 
\begin{equation}
B_{+-}(\Delta \eta) = \frac{1}{2}\Big(C_{+-}(\Delta \eta) + C_{-+}(\Delta \eta) -
C_{--}(\Delta \eta) - C_{++}(\Delta \eta)\Big).
\label{Eq:BFDefinitionInEta}
\end{equation}

Each term of Eq.~\ref{Eq:BFDefinitionInEta}, is corrected for detector and tracking inefficiencies as well as for 
acceptance effects and can be written as $C_{ab} = (N_{ab}/N_b)/f_{ab}$. The factors $f_{ab}$ (where in the case of charged particles, $a$ and $b$ correspond to the charge i.e. $f_{+-}$, $f_{-+}$, $f_{++}$ and $f_{--}$) represent the probability that given a particle $a$ is reconstructed, a second particle  emitted at a relative pseudorapidity or azimuthal angle (\deta~or \dphi, respectively), would also be detected. These terms are defined as the product of the single particle 
tracking efficiency $\varepsilon(\eta,\varphi,p_{\rm T})$ and the acceptance term $\alpha(\Delta\eta,\Delta\varphi)$. The way they are extracted in this analysis with a data driven method is described in one of the following sections.

For a neutral system, every charge has an opposite 
balancing partner and the balance function would integrate to unity. However, this normalization does not hold if not all charged particles are included in the calculation due to specific momentum range or particle type selection.


The width of the balance function distribution can be used to quantify 
how tightly the balancing charges are correlated. It can be characterized by the average 
\wdeta~or \wdphi~in case of studies in pseudorapidity 
or the azimuthal angle, respectively. The mathematical expression for the case of correlations in
pseudorapidity is given in Eq.~\ref{Eq:WeightedAverage} (similar for $\langle \Delta \varphi \rangle$)
\begin{equation}
\langle \Delta \eta \rangle = \sum_{i=1}^k{[B_{+-}(\Delta \eta _i) \cdot \Delta \eta _i]}/\sum_{i=1}^k{B_{+-}(\Delta \eta _i)},
\label{Eq:WeightedAverage}
\end{equation}
\noindent where $B_{+-}(\Delta \eta _i)$ is the balance function value for each bin $\Delta \eta_i$, with the sum running over all bins $k$.

Experimentally, the balance function for non-identified particles was studied by the STAR 
Collaboration in Au--Au collisions at $\sqrt{s_{\rm NN}} = 130$~GeV \cite{Ref:BalanceFunctionSTAR1}, 
followed by the NA49 experiment in Pb--Pb collisions at the highest SPS energy \cite{Ref:BalanceFunctionNA49}. 
Both experiments reported the narrowing of the balance function in \deta~in more central compared to peripheral collisions. The results were qualitatively in agreement with theoretical expectations for a system with a long-lived QGP phase and exhibiting delayed hadronization. These results 
triggered an intense theoretical investigation of their interpretation \cite{Ref:PrattPhysRevC68, Ref:PrattPhysRevC69, Ref:Florkowski, Ref:Bozek, Ref:Voloshin, Ref:Rafelski, Ref:Bialas}. In \cite{Ref:PrattPhysRevC68}, it was suggested that the balance function could be distorted by the excess of positive charges due to the protons of the incoming beams (unbalanced charges). This effect is expected to be reduced at higher collision energy, leaving a system at mid-rapidity that is net-baryon free. Also in \cite{Ref:PrattPhysRevC68}, it was proposed to perform balance function studies in terms of the relative invariant momentum of the particle pair, to eliminate the sensitivity to collective flow. In \cite{Ref:PrattPhysRevC69}, it was shown that purely hadronic models predict a modest broadening of the balance function for central heavy-ion collisions, contrary to the experimentally measured narrowing. It was also shown that thermal models were in agreement with the (at that time) published data, concluding that charge conservation is local at freeze-out, consistent with the delayed charged-creation scenario \cite{Ref:PrattPhysRevC69}. Similar agreement with the STAR data was reported in \cite{Ref:Florkowski}, where a thermal model that included resonances was used. In \cite{Ref:Bozek}, the author showed that the balance function, when measured in terms of the relative azimuthal angle of the pair, is a sensitive probe of the system's collective motion and in particular of its radial flow. In \cite{Ref:Voloshin}, it was suggested that radial flow is also the driving force of the narrowing of the balance function in pseudorapidity, with its width being inversely proportional to the transverse mass, $m_{\rm{T}} = \sqrt{m^2 + p_{\rm{T}}^2}$. In parallel in \cite{Ref:Rafelski, Ref:Bialas}, the authors attributed the narrowing of the balance function for more central collisions to short range correlations in the QGP at freeze-out.

Recently, the STAR Collaboration 
extended their balance function studies in Au--Au collisions at $\sqrt{s_{\rm NN}} = 200$~GeV \cite{Ref:BalanceFunctionSTAR2}, confirming the strong centrality dependence of the width in \deta~but also revealing a similar dependence in \dphi, the latter being mainly attributed to radial flow. 
Finally, in \cite{Ref:PrattAzimuthalAngle} the authors fitted the experimentally measured balance function at the top RHIC energies with a blast-wave parameterization and argued that in \dphi~the results could be explained by larger radial flow in more central collisions. However the results in \deta~could only be reproduced when considering the separation of charges at freeze-out implemented in the model. They also stressed the importance of performing a multi-dimensional analysis. In particular, they presented how the balance function measured with respect to the orientation of the reaction plane (i.e. the plane of symmetry of a collision defined by the impact parameter vector and the beam direction) could probe potentially one of the largest sources of background in studies related to parity violating effects in heavy-ion collisions \cite{Ref:AliceParity}. 

In this Letter we report the first results of the balance function measurements in Pb--Pb collisions at 
$\sqrt{s_{\rm NN}} = 2.76$~TeV with the ALICE detector \cite{Ref:AlicePPR,Ref:AliceJinst}. The Letter is organized as follows: 
Section~\ref{Section:ExpSetup} briefly describes the experimental setup, while details about 
the data analysis are presented in Section~\ref{Section:DataAnalysis}. In Section~\ref{Section:Results} 
we discuss the main results followed by a detailed comparison with different models in 
Section~\ref{Section:Comparison}. In the same section we present the energy dependence of the 
balance function. We conclude with the 
summary and a short outlook. 

\section{Experimental setup}
\label{Section:ExpSetup}

ALICE \cite{Ref:AliceJinst} is the dedicated heavy-ion detector at the LHC, designed 
to cope with the high charged-particle densities measured in central Pb--Pb collisions \cite{Ref:AlicedNdeta}. 
The experiment consists of a large number of detector subsystems inside a solenoidal magnet (0.5~T). 
The central tracking systems of ALICE provide full azimuthal coverage within a pseudorapidity window 
$|\eta| < 0.9$. They are also optimized to provide good momentum resolution ($\approx 1\%$ at \pt~$< 1$~GeV/$c$) 
and particle identification (PID) over a broad momentum range, the latter being important for the future, particle type dependent balance function studies.

For this analysis, the charged particles were reconstructed using the \textit{Time Projection Chamber} (\TPC) 
\cite{Ref:ALICETPC}, which is the main tracking detector of the central barrel. In addition, a complementary analysis relying on the combined tracking of the \TPC~and the \textit{Inner Tracking System} (\ITS) was performed. The \ITS~consists 
of six layers of silicon detectors employing three different technologies. The two innermost layers are \textit{Silicon Pixel Detectors} (\SPD), followed by two layers of \textit{Silicon Drift Detectors} (\SDD). Finally the two outermost layers are double-sided \textit{Silicon Strip Detectors} (\SSD).

The position of the primary interaction was determined by the \TPC~and by the \SPD, depending 
on the tracking mode used. A set of forward detectors, namely the \VZERO~scintillator arrays, were used in the trigger logic and also for the centrality 
determination \cite{Ref:ALICECentrality}. The \VZERO~detector consists of two arrays of scintillator counters, the \VZERO-A and 
the \VZERO-C, positioned on each side of the interaction point. They cover the pseudorapidity 
ranges of $2.8 < \eta < 5.1$ and $-3.7 < \eta < -1.7$ for \VZERO-A and \VZERO-C, respectively. 

For more details on the ALICE experimental setup, see \cite{Ref:AliceJinst}.

\section{Data analysis}
\label{Section:DataAnalysis}

Approximately 15~$\times$~$10^6$ Pb--Pb events, recorded during the first 
LHC heavy-ion run in 2010 at $\sqrt{s_{\rm NN}} = 2.76$~TeV, were analyzed. A minimum bias trigger was used, requiring two pixel chips hit in the \SPD~in coincidence with a 
signal in the \VZERO-A and \VZERO-C detectors. Measurements were also made with the requirement changed 
to a coincidence between signals from the two sides of the \VZERO~detectors. An offline event 
selection was also applied in order to reduce the contamination from background events, such as electromagnetic and beam-gas interactions. All events were required to have a reconstructed vertex position along the beam axis ($V_z$) with $\left| V_z \right| < 10$~cm from the nominal interaction point.

\begin{figure}[thb!]
\includegraphics[width=\linewidth]{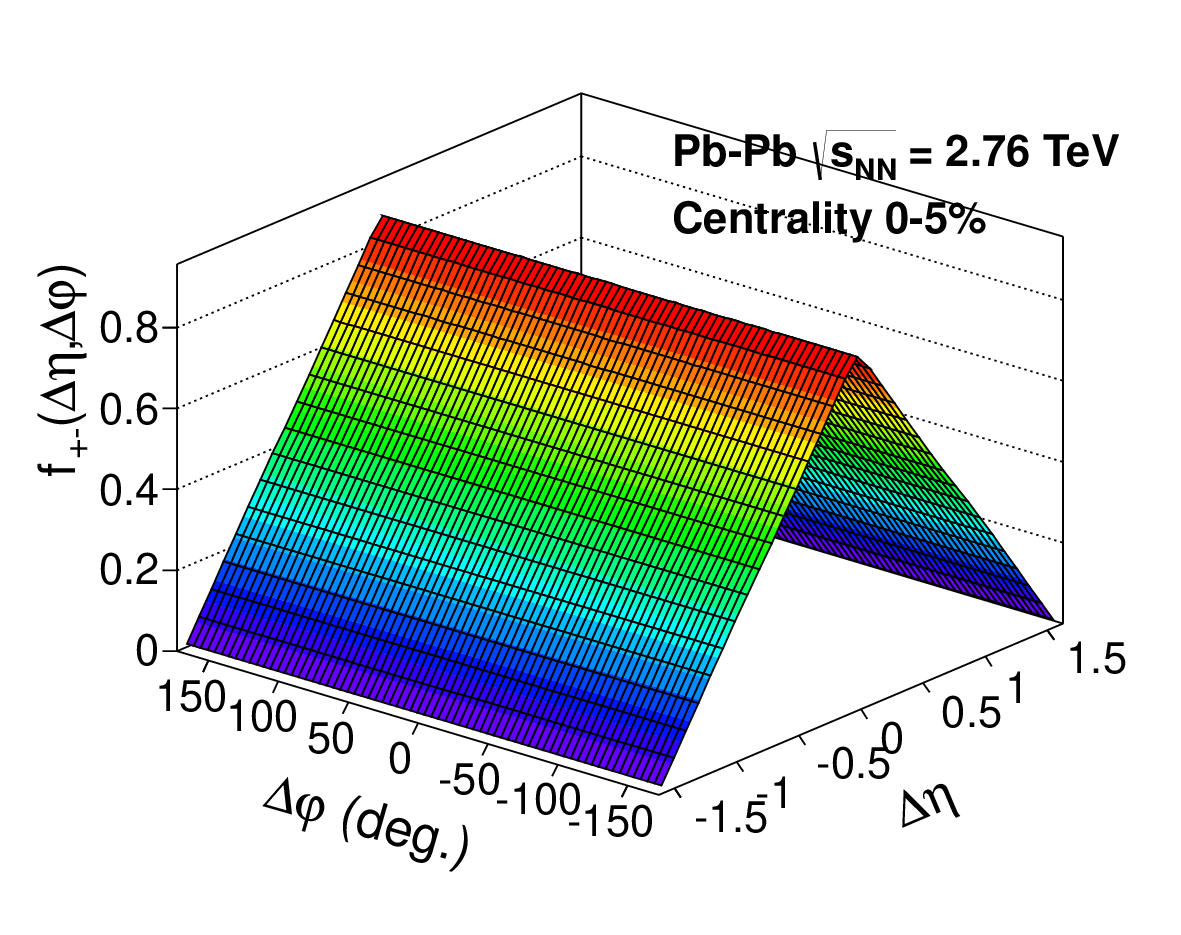}
\caption{(Color online). The correction factor $f_{+-}(\Delta\eta,\Delta\varphi)$ for the 5$\%$ most central Pb--Pb 
collisions, extracted from the single particle tracking efficiencies $\varepsilon(\eta,\varphi,p_{\rm T})$ and the acceptance terms $\alpha(\Delta\eta,\Delta\varphi)$ (see text for details).}
\label{fig:AcceptanceComparison}
\end{figure}

\begin{figure*}[tb]
\includegraphics[width=\linewidth]{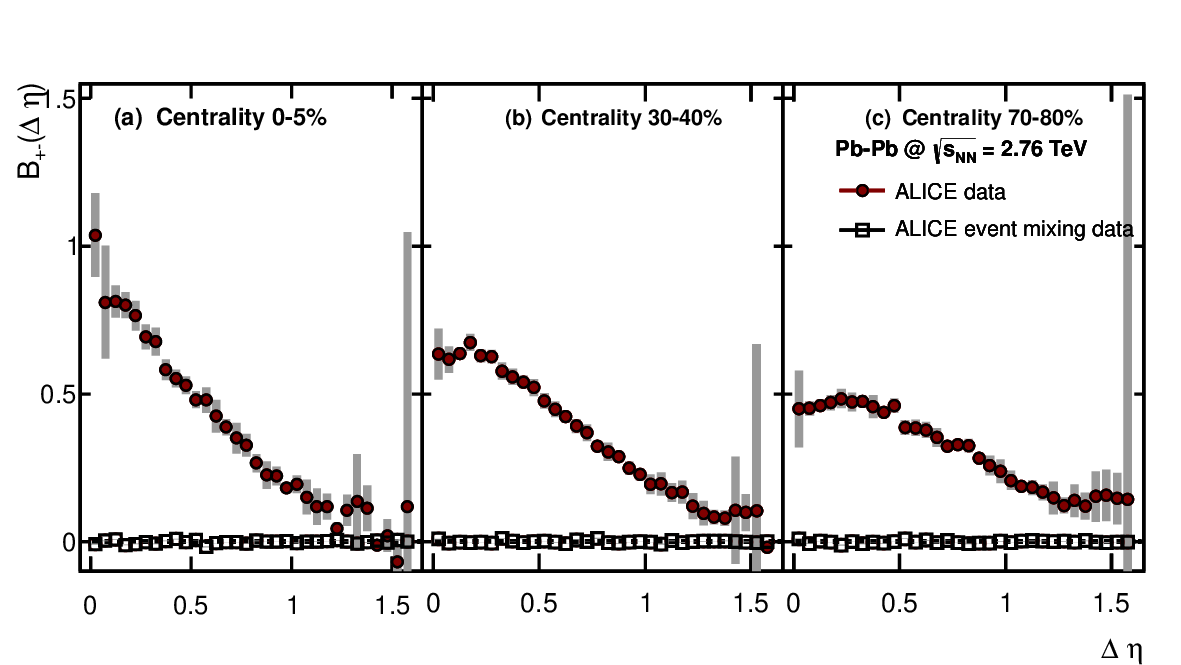}
\caption{(Color online). Balance function as a function of \deta~for different centrality classes: 0-5$\%$ (a), 
30-40$\%$ (b) and 70-80$\%$ (c).  Mixed events results, not corrected for the detector effects, are shown by open squares. See text for details.}
\label{fig:bfDistributionsAliceInDeltaEta}
\end{figure*}

The data were sorted according to centrality classes, reflecting the geometry of the collision (i.e. impact 
parameter),  which span $0-80\%$ of the inelastic cross section. The $0-5\%$ bin corresponds to the most 
central (i.e. small impact parameter) and the $70-80\%$ class to the most peripheral (i.e. large impact parameter) collisions. The centrality of the collision was estimated using the charged particle  
multiplicity distribution and the distribution of signals from the \VZERO~scintillator detectors. Fitting these distributions with a Glauber model \cite{Ref:Glauber}, the centrality classes are mapped to the corresponding mean number of participating nucleons $\langle N_{\mathrm{part}} \rangle$ \cite{Ref:NPart}. Different centrality estimators (i.e. \TPC~tracks, \SPD~clusters) were used to investigate the systematic uncertainties. Further details on the centrality determination can be found in \cite{Ref:ALICECentrality}.

To select charged particles with high efficiency and to minimize the contribution from background tracks 
(i.e. secondary particles originating either from weak decays or from the interaction of particles with the 
material), all selected tracks were required to have at least 70 reconstructed space points out of the 
maximum of 159 possible in the \TPC. The $\langle \chi^2 \rangle$ per degree of freedom  the momentum fit was required to be below $2$. To further reduce the contamination from background 
tracks, a cut on the distance of closest approach between the tracks and the primary vertex ($dca$) 
was applied $(dca_{xy}/d_{xy})^2 + (dca_{z}/d_{z})^2 < 1$ with $d_{xy} = 2.4$~cm and $d_{z} = 3.2$~cm. In the parallel analysis, with the combined tracking of the \TPC~and the \ITS, the values of $d_{xy} = 0.3$~cm and $d_{z} = 0.3$~cm were used, profiting from the better $dca$ resolution that the \ITS~provides.
Finally, we report the results for the region of $|\eta| < 0.8$ and $0.3 < p_{\rm{T}} < 1.5$~GeV/$c$. The $p_{\rm{T}}$ range is chosen to ensure a high tracking efficiency (lower cut) and a minimum contribution from (mini-)jet correlations (upper cut).

\begin{figure*}[tb]
\includegraphics[width=\linewidth]{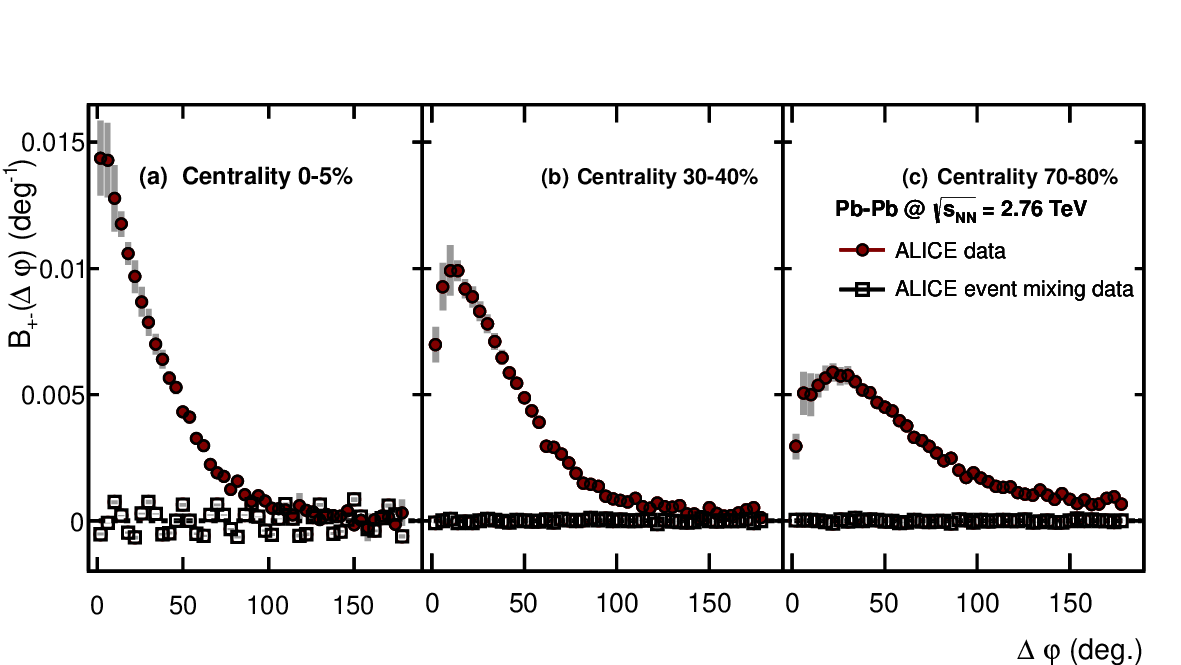}
\caption{(Color online).  Balance function as a function of \dphi~for different centrality classes: 0-5$\%$ (a), 
30-40$\%$ (b) and 70-80$\%$ (c).  Mixed events results, not corrected for the detector effects, are shown by open squares. See text for details.}
\label{fig:bfDistributionsAliceInDeltaPhi}
\end{figure*}

\section{Results}
\label{Section:Results}

As discussed in the introduction, the correction factors $f_{+-}$, $f_{-+}$, $f_{++}$, and $f_{--}$ are needed to eliminate the dependence of the balance function on the detector acceptance and tracking inefficiencies.
The tracking efficiency is extracted from a detailed Monte Carlo simulation of the ALICE detector based on GEANT3 \cite{Ref:GEANT}.It depends on the particle's transverse momentum, rising steeply from $0.2$ up to $0.5$~GeV/$c$, where it reaches the saturation value of $85\%$. The acceptance part of the correction factors, $\alpha(\Delta\eta,\Delta\varphi)$, is extracted from mixed events. The mixed events are generated by taking all two-particle non-same-event combinations for a collection of a few ($\approx$ 5) events with similar values of the z position of the reconstructed vertex ($|\Delta V_z| < 5$~cm). 
In addition, the events used for the event mixing belonged to the same centrality class and had multiplicities 
that did not differ by more than 1-2$\%$, depending on the centrality.
Figure~\ref{fig:AcceptanceComparison} presents the correction factor for the distribution of pairs of particles 
with opposite charge as a function of the relative pseudorapidity and azimuthal angle differences 
for the 5$\%$ most central Pb--Pb collisions. The maximum value 
is observed for $\Delta\eta = 0$ and is equal to the \pt-integrated single particle efficiency. The distribution decreases to $\approx 0$ near the edge of the acceptance i.e. $|\Delta\eta| \approx 1.6$. This reduction reflects 
the decrease of the probability of detecting both balancing charges as the relative pseudorapidity difference 
increases. The correction factor is constant as a function of \dphi. 

The measured balance function is averaged over positive and negative values of $\Delta \eta$ ($\Delta \varphi$) and reported only for positive values. The integrals of the balance function over the reported region are close to 0.5, reflecting the fact that most of the balancing charges are distributed in the measured region.

Figure~\ref{fig:bfDistributionsAliceInDeltaEta} presents the balance functions as a function of the 
relative pseudorapidity \deta~for three different centrality classes: the 0-5$\%$ (most central), 
the 30-40$\%$ (mid--central) and the 70-80$\%$ (most peripheral) centrality bins. It is seen that the balance function, in full circles, gets narrower for more central collisions.  Figure~\ref{fig:bfDistributionsAliceInDeltaEta} presents also the balance functions for mixed events, not corrected for detector effects, represented by the open squares. These balance functions, 
fluctuate around zero as expected for a totally uncorrelated 
sample where the charge is not conserved. 

Figure~\ref{fig:bfDistributionsAliceInDeltaPhi} presents the balance functions as a function of the relative 
azimuthal angle for the same centrality classes as in Fig.~\ref{fig:bfDistributionsAliceInDeltaEta}. 
The balance functions calculated using mixed events and not corrected for the tracking efficiency exhibit a distinct modulation originating 
from the 18 sectors of the \TPC. This modulation is more pronounced for more central collisions, since 
the charge dependent acceptance differences scale with multiplicity. The efficiency-corrected balance functions, represented by the full markers, indicate that these detector effects are 
successfully removed. 
Narrowing of the balance function in more central events has been also observed in this representation. A decrease of the balance function at small \dphi~(i.e. for $\Delta \varphi \leq 10^\circ$) can be 
observed for the mid-central and peripheral collisions. This can be attributed to short-range correlations between pairs of same and opposite charge, 
such as HBT and Coulomb effects \cite{Ref:PrattAzimuthalAngle}.


In both Fig.~\ref{fig:bfDistributionsAliceInDeltaEta} and Fig.~\ref{fig:bfDistributionsAliceInDeltaPhi} 
as well as in the next figures, the error bar of each point corresponds to the statistical uncertainty (typically the size of the marker). The systematic uncertainty is represented by the shaded band around each point. The origin 
and the value of the assigned systematic uncertainty on the width of the balance function, calculated for each centrality and for both \deta~and \dphi, will be 
discussed in the next paragraph. 

The data sample was analyzed separately for two magnetic field configurations. The two data samples had comparable statistics. The 
maximum value of the systematic uncertainty, defined as half of the difference between the balance functions in these two cases, 
is found to be less than $1.3\%$ over all centralities. In addition, we estimated the 
contribution to the systematic uncertainty originating from the centrality selection, by determining the 
centrality not only with the \VZERO~detector but alternatively using the multiplicity of the \TPC~tracks or the number of clusters of the second \SPD~layer. This resulted in an additional maximum contribution 
to the estimated systematic uncertainty of $0.8 \%$ over all centralities. Furthermore, we investigated the influence of the 
ranges of the cuts in parameters such as the position of the primary vertex in the $z$ coordinate ($\left| V_z \right| < 6-12$~cm), the 
$dca$ ($d_{xy} < 1.8-2.4$~cm and $d_{z} < 2.6-3.2$~cm), and the number of required \TPC~clusters ($N_{clusters} (TPC) > 60-90$). This was done by varying the relevant 
ranges, one at a time, and again assigning half of the difference between the lower and higher value of 
the width to the systematic uncertainty. The maximum contribution from these sources was estimated to be 
$1.3 \%$, $1.1 \%$ and $1.3 \%$ for the three parameters, respectively. We also studied the influence 
of the different tracking modes used by repeating the analysis using tracks reconstructed by the combination 
of the \TPC~and the \ITS~(global tracking). The resulting maximum contribution to the systematic uncertainty 
of the width from this source is $1.1 \%$, again over all centralities. Finally, the applied acceptance corrections result in large fluctuations of the balance function points for some centralities towards the edge of the acceptance (i.e. large values of $\Delta\eta$), which originates from the division of two small numbers. To account for this, we average over several bins at these high values of $\Delta\eta$ to extract the weighted average. This procedure results in an uncertainty that has a maximum value of 5$\%$ over all centralities. All these contributions are summarized in Table~\ref{tab:systematic}. The final 
systematic uncertainty for each centrality bin was calculated by adding all the different sources in quadrature. The resulting values for the 0-5$\%$, 30-40$\%$ and 70-80$\%$ centrality bins were estimated to be 
2.5$\%$, 3.0$\%$ and 3.6$\%$, respectively, in $\langle \Delta\eta \rangle$ (1.9$\%$, 1.2$\%$ and 2.4$\%$, respectively, in $\langle \Delta\varphi \rangle$).

\begin{table}[ht]
\centering
\caption{The maximum value of the systematic uncertainties on the width of the balance function over 
all centralities for each of the sources studied.}
\begin{tabular}{c c c}
\hline
\multicolumn{3}{|c|}{Systematic Uncertainty} \\
\hline
\hline
Category & Source & Value (max) \\ 
\hline

Magnetic field & (++)/(- -) & $1.3 \%$ \\
\hline
Centrality estimator & \VZERO, \TPC, \SPD & $0.8 \%$ \\
\hline
                        & $dca$ & $1.3 \%$ \\
Cut variation & $N_{clusters} (TPC)$ & $1.1 \%$ \\
                        & $\Delta V_{z}$ & $1.3 \%$ \\
\hline
Tracking & \TPC, Global & $1.1 \%$ \\ 
\hline
Binning & Extrapolation to large $\Delta\eta$ & $5.0 \%$ \\ 
\hline
\end{tabular}
\label{tab:systematic} 
\end{table}

\section{Discussion}
\label{Section:Comparison}

\subsection{Centrality dependence}
\label{Section:ComparisonModels}

\begin{figure}[thb!]
\centering
\includegraphics[width=0.92\linewidth]{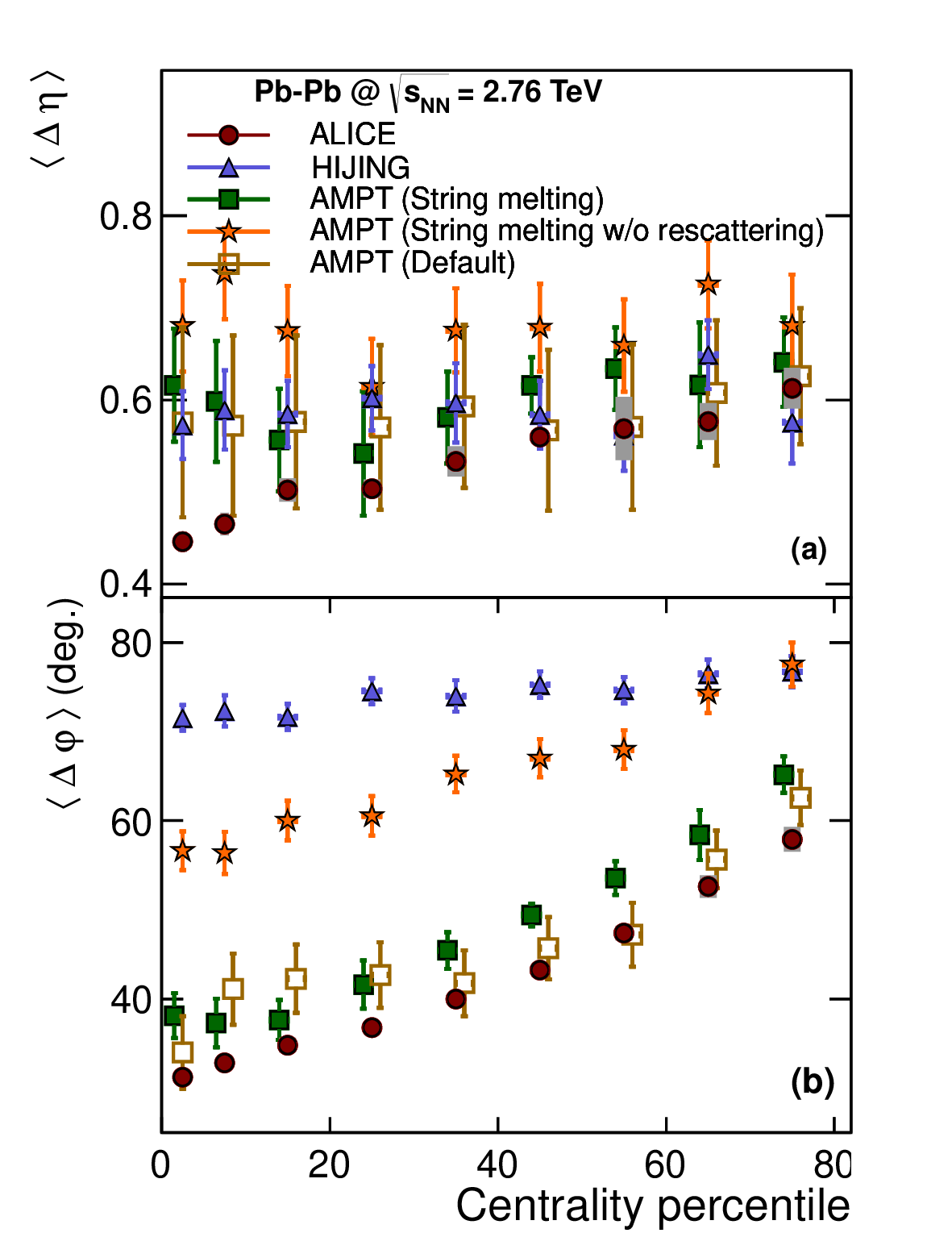}
\caption{(Color online).  The centrality dependence of the width of the balance function \wdeta~
and $\langle \Delta \varphi \rangle$, for the correlations studied in terms of the relative pseudorapidity (a) and 
the relative azimuthal angle (b), respectively. The data points are compared to the predictions from HIJING \cite{Ref:Hijing}, and AMPT \cite{Ref:Ampt}. }
\label{fig:ModelComparison}
\end{figure}

The width  of the balance function (Eq.~\ref{Eq:WeightedAverage}) as a function of the centrality percentile
is presented in Fig.~\ref{fig:ModelComparison}. Central (peripheral) collisions correspond to small (large) centrality percentile. The width 
is calculated in the entire interval where the balance function was measured (i.e. $0.0 < \Delta \eta < 1.6$ and $0^o < \Delta \varphi < 180^o$). Both results in terms of correlations in the relative pseudorapidity ($\langle \Delta \eta \rangle$, upper 
panel, Fig.~\ref{fig:ModelComparison}-a) and the relative azimuthal angle ($\langle \Delta \varphi \rangle$, lower panel, Fig.~\ref{fig:ModelComparison}-b) are shown. The experimental data 
points, represented by the full red circles, exhibit a strong centrality dependence: 
more central collisions correspond to narrower distributions (i.e. moving from right to left along the x-axis) for both \deta~and \dphi. Our results are compared to 
different model predictions, such as HIJING \cite{Ref:Hijing} and different versions of a multi-phase transport 
model (AMPT) \cite{Ref:Ampt}. The error bars in the results from these models represent the statistical uncertainties.

The points from the analysis of HIJING Pb--Pb events at $\sqrt{s_{\rm NN}} = 2.76$~TeV, represented by the 
blue triangles, show little centrality dependence in both projections. The slightly narrower balance functions for central collisions might be related to the fact that HIJING is not just a simple superposition of single pp collisions; jet-like effects as well as increased resonance yields in central collisions could be reflected as additional correlations. The balance function widths generated by HIJING are much larger than those measured in the data, consistent with the fact that the model lacks collective flow.

\begin{figure}[thb!]
\centering
\includegraphics[width=0.77\linewidth]{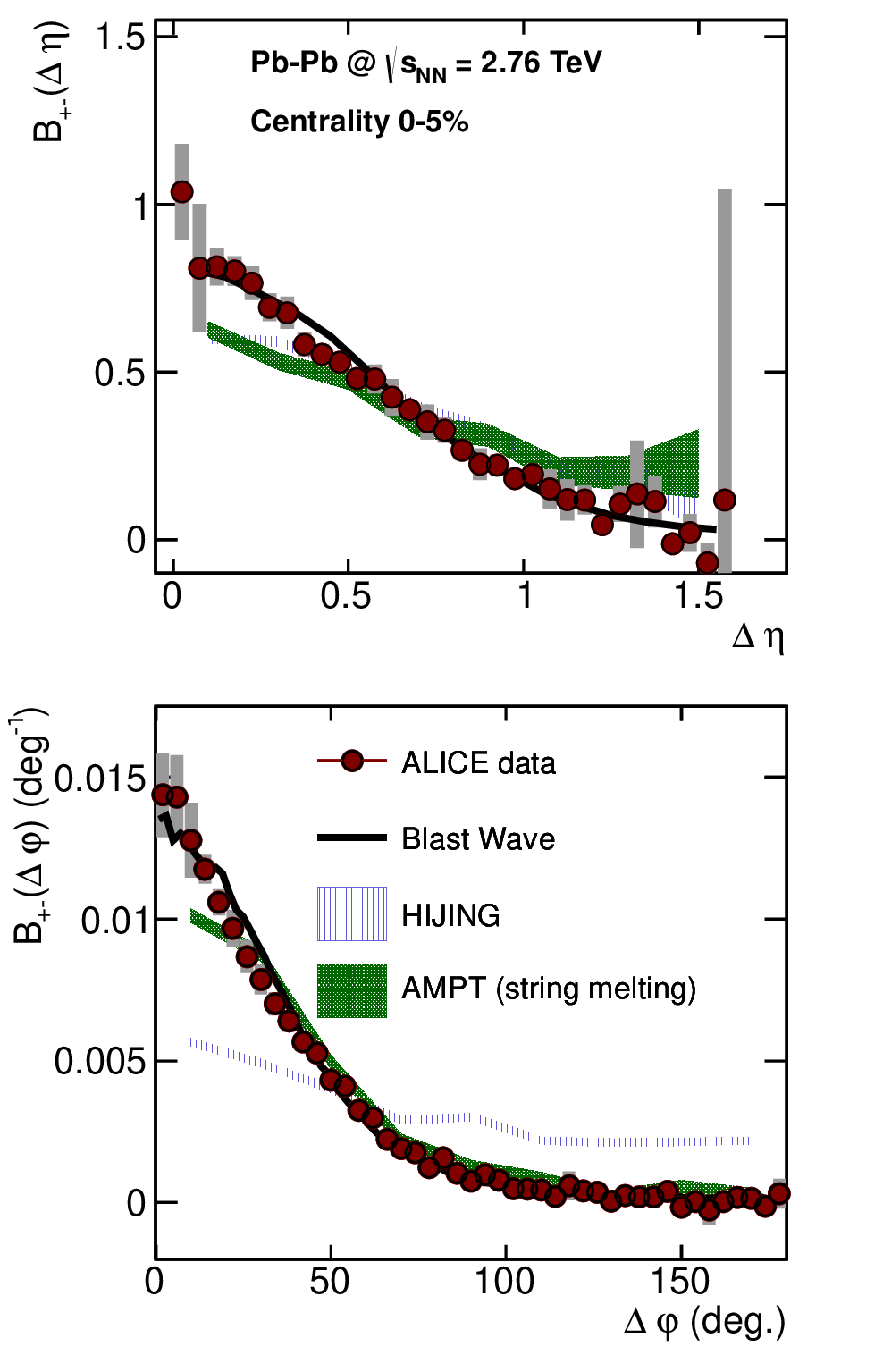}
\caption{(Color online). The balance functions for the 5$\%$ most central Pb--Pb collisions measured by ALICE 
as a function of the relative pseudorapidity (a) and the relative azimuthal angle (b). The experimental points 
are compared to predictions from HIJING \cite{Ref:Hijing}, AMPT \cite{Ref:Ampt} and from a thermal blast 
wave \cite{Ref:BlastWave,Ref:BlastWave2}.}
\label{fig:BlastWaveComparison}
\end{figure}

In addition, we compare our data points to the results from the analysis of events from three different 
versions of AMPT in Fig.~\ref{fig:ModelComparison}. The AMPT model consists of two different 
configurations: the \emph{default} and the \emph{string melting}. Both are based on HIJING to describe 
the initial conditions. The partonic evolution is described by the Zhang's parton cascade (ZPC) \cite{Ref:ZPC}. In the 
\emph{default} AMPT model, partons are recombined with their parent strings when they stop interacting, 
and the resulting strings are converted to hadrons using the Lund string fragmentation model. In the 
\emph{string melting} configuration a quark coalescence model is used instead to combine partons 
into hadrons. The final part of the whole process, common between the two configurations,  
consists of the hadronic rescattering which also includes the decay of resonances. 

The filled green squares represent the results of the analysis of the \emph{string melting} AMPT events 
with parameters tuned \cite{Ref:AmptTune}  to reproduce the measured elliptic flow (v$_2$) values of 
non-identified particles at the LHC \cite{Ref:AliceFlow}. The width of the balance functions when studied 
in terms of the relative pseudorapidity exhibit little centrality dependence despite the fact 
that the produced system exhibits significant collective behavior \cite{Ref:AmptTune}. 
However, the width of the balance function in \dphi~is in 
qualitative agreement with the centrality dependence of the experimental points. This is consistent with the expectation that the balance function when studied as a function of \dphi~can be used as a measure of 
radial flow of the system, as suggested in \cite{Ref:Bozek,Ref:PrattAzimuthalAngle}. We also studied the 
same AMPT configuration, i.e. the \emph{string melting}, this time switching off the last part where the 
hadronic rescattering takes place, without altering the decay of resonances. The resulting points, indicated 
with the orange filled stars in Fig.~\ref{fig:ModelComparison}, demonstrate a similar qualitative behavior 
as in the previous case: no centrality dependence of \wdeta~and a significant decrease 
of \wdphi~for central collisions. On a quantitative level 
though, the widths in both projections are larger than the ones obtained in the case where hadronic rescattering 
is included. This can be explained by the fact that within this model, a significant part of radial flow of the 
system is built during this very last stage of the system's evolution. Therefore, the results are consistent with 
the picture of having the balancing charges more focused under the influence of this collective motion, 
which is reflected in a narrower balance function distribution. In addition, we analyzed AMPT events produced 
using the \emph{default} configuration, which results in smaller v$_n$ flow coefficients but harder spectra 
than the \emph{string melting}. The extracted widths of the balance functions are represented by the open 
brown squares and exhibit similar behavior as the results from the \emph{string melting} configuration. In 
particular, the width in \deta~shows little centrality dependence while the values are in agreement 
with the ones calculated from the \emph{string melting}. The width in \dphi~shows similar (within the 
statistical uncertainties) quantitative centrality dependence as the experimental data points. This latter effect is consistent with the observation of having a system exhibiting larger radial flow 
with the \emph{default} version.\footnote{We recently confirmed that AMPT does not conserve the charge. The influence of this effect to our measurement cannot be easily quantified. However we still consider interesting and worthwhile to point out that this model describes in a qualitative (and to some extent quantitative) way the centrality dependence of $\langle \Delta \varphi \rangle$.}

Finally, we fit the experimentally measured values with a thermal blast-wave model \cite{Ref:BlastWave,Ref:BlastWave2}. 
This model, assumes that the radial expansion velocity is proportional to the distance from the center of the system and takes into account the resonance production and decay. It also incorporates the local charge conservation, by generating ensembles of particles with zero total charge. Each particle of an ensemble is emitted by a fluid element with a common collective velocity 
following the single-particle blast-wave parameterization with the additional constraint of being emitted 
with a separation at kinetic freeze-out from the neighboring particle sampled from a Gaussian with 
a width denoted as $\sigma_{\eta}$ and $\sigma_{\varphi}$ in the pseudorapidity space and the azimuthal 
angle, respectively. The procedure that we followed started from tuning the input parameters of the model to 
match the average \pt~values extracted from the analysis of identified particle spectra \cite{Ref:AliceIdentifiedSpectra} 
as well as the v$_2$ values for non-identified particles reported by ALICE \cite{Ref:AliceFlow}. We then 
adjust the widths of the parameters $\sigma_{\eta}$ and $\sigma_{\varphi}$ to match the experimentally 
measured widths of the balance function, \wdeta~and $\langle \Delta \varphi \rangle$. 
The resulting values of $\sigma_{\eta}$ and $\sigma_{\varphi}$ are listed in Table~\ref{tab:blastwave}. We 
find that $\sigma_{\eta}$ starts from $0.28 \pm 0.05$ for the most central Pb--Pb collisions reaching 
$0.52 \pm 0.07$ for the most peripheral, while $\sigma_{\varphi}$ starts from $0.30 \pm 0.10$ evolving to $0.76 \pm 0.01$ for the 60-70$\%$ centrality bin. 

\begin{table}[ht]
\caption{The values of $\sigma_{\eta}$ and $\sigma_{\varphi}$ extracted by fitting the centrality dependence 
of both \wdeta~and \wdphi~with the blast-wave parameterization 
of \cite{Ref:BlastWave,Ref:BlastWave2}. }
\centering
\begin{tabular}{c c c}
\hline
\multicolumn{3}{|c|}{Results from the fit with the blast-wave model} \\
\hline
\hline
Centrality & $\sigma_{\eta}$ & $\sigma_{\varphi}$ \\ 
\hline
0-5$\%$ &  $0.28 \pm 0.05$ & $0.30 \pm 0.10$ \\
\hline
5-10$\%$ & $0.32 \pm 0.05$ & $0.35 \pm 0.07$ \\
\hline
10-20$\%$ & $0.31 \pm 0.05$ & $0.36 \pm 0.08$ \\
\hline
20-30$\%$ & $0.36 \pm 0.03$ & $0.43 \pm 0.05$ \\
\hline
30-40$\%$ & $0.43 \pm 0.04$ & $0.52 \pm 0.05$ \\
\hline
40-50$\%$ & $0.42 \pm 0.04$ & $0.54 \pm 0.06$ \\
\hline
50-60$\%$ & $0.44 \pm 0.07$ & $0.64 \pm 0.06$ \\
\hline
60-70$\%$ & $0.52 \pm 0.07$ & $0.76 \pm 0.01$ \\
\hline
\end{tabular}
\label{tab:blastwave} 
\end{table}

Figure~\ref{fig:BlastWaveComparison} presents the detailed comparison of the model results with the measured balance functions as a function of \deta~(a) and \dphi~(b) for the 5$\%$ most central 
Pb--Pb collisions. The data points are represented by the full markers and are compared with HIJING (dashed black line), AMPT \emph{string melting} (full green line) and the thermal blast-wave (full black line). The 
distributions for HIJING and AMPT are normalized to the same integral to facilitate the direct comparison of 
the shapes and the widths. It is seen that for correlations in the relative pseudorapidity, both HIJING and AMPT 
result in similarly wider distributions. As mentioned before, the blast-wave model is tuned to reproduce 
the experimental points, so it is not surprising that the relevant curve not only reproduces the same 
narrow distribution but describes fairly well also its shape. For the correlations in \dphi~the 
HIJING curve clearly results in a wider balance function distribution. On the other hand, there is a very 
good agreement between the AMPT curve and the measured points, with the exception of the first bins 
(i.e. small relative azimuthal angles) where the magnitude of $B_{+-}(\Delta \varphi$) is significantly larger in real 
data. This suggests that there are additional correlations present in these small ranges of \dphi~
in data than what the model predicts.

\subsection{Energy dependence}
\begin{figure}[thb!]
\centering
\includegraphics[width=0.92\linewidth]{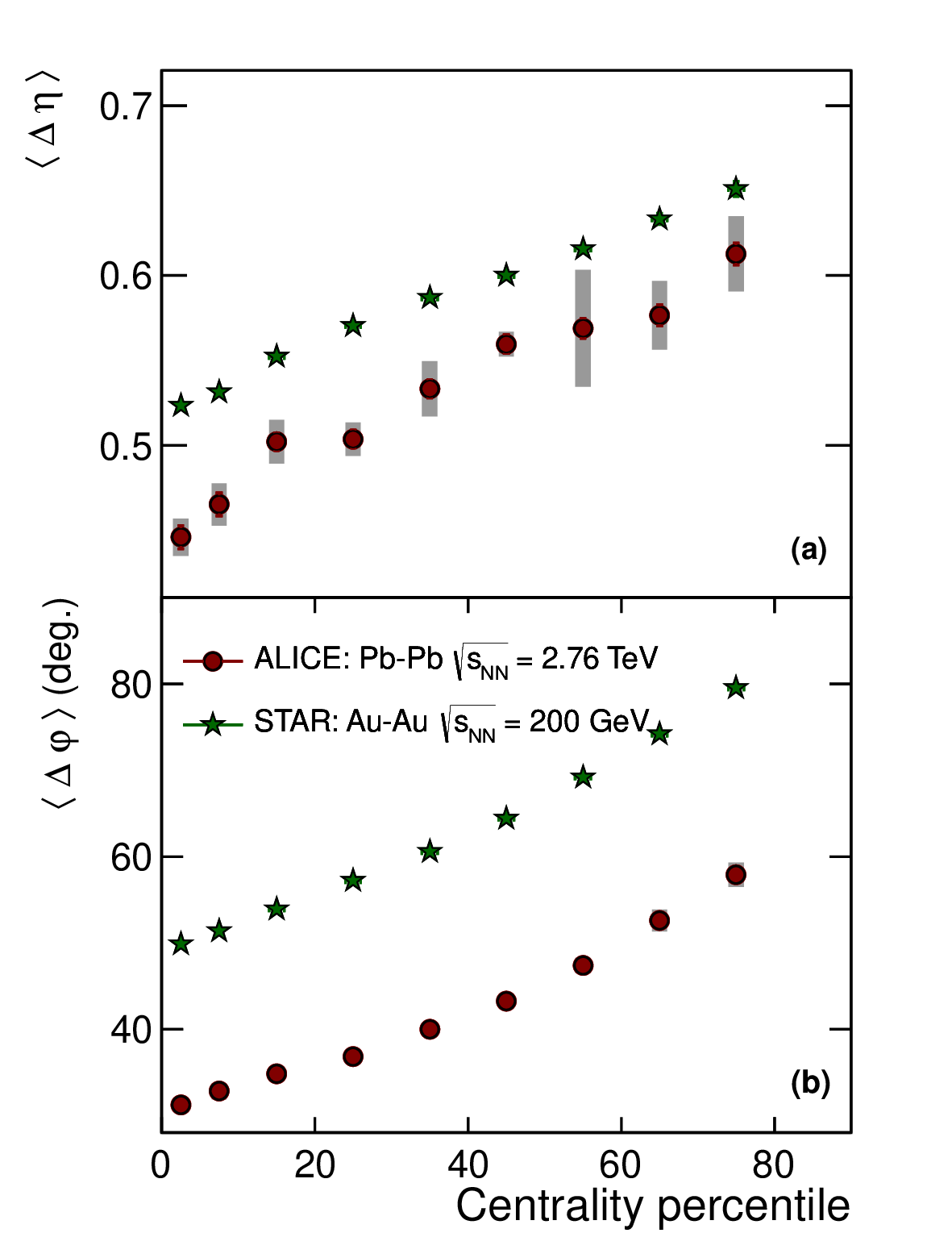}
\caption{(Color online). The centrality dependence of the balance function width $\langle \Delta \eta \rangle$ 
(a) and $\langle \Delta \varphi \rangle$ (b). The ALICE points are compared to results from STAR \cite{Ref:BalanceFunctionSTAR2}. The STAR results have been corrected for the finite acceptance as suggested in \cite{Ref:PrattPhysRevC65}.}
\label{fig:ExperimentComparison}
\end{figure}

Figure~\ref{fig:ExperimentComparison} presents the comparison of our results for the centrality dependence (i.e. as a function of the centrality percentile) of the width 
of the balance function, $\langle \Delta \eta \rangle$ (Fig.~\ref{fig:ExperimentComparison}-a) and $\langle \Delta \varphi \rangle$ (Fig.~\ref{fig:ExperimentComparison}-b), with results from STAR \cite{Ref:BalanceFunctionSTAR2} in Au--Au collisions at $\sqrt{s_{\rm NN}} =$~200~GeV (stars). 
The ALICE points have been corrected for acceptance and detector effects, using the correction factors $f_{ab}$, discussed in the introduction. To make a proper comparison with the STAR measurement, where such a correction was not applied, we employ the procedure suggested in \cite{Ref:PrattPhysRevC65} to the RHIC points. Based on the assumption of a boost-invariant system the balance function studied in a given pseudorapidity window $B_{+-}(\Delta \eta|\eta_{\mathrm{max}}$) can be related to the balance function for an infinite interval according to the formula of Eq.~\ref{Eq:AcceptanceCorrection}

\begin{center}
\begin{equation}
B_{+-}(\Delta \eta|\eta_{\mathrm{max}}) = B_{+-}(\Delta \eta|\infty) \cdot \Big(1 - \frac{\Delta \eta}{\eta_{\mathrm{max}}}\Big).
\label{Eq:AcceptanceCorrection}
\end{equation}
\end{center}

\noindent This procedure results in similar corrections as to the case where the $f_{ab}$ are used, if the acceptance is flat in $\eta$ (which is a reasonable assumption for the acceptance of STAR).\footnote{We do not compare our results to the data from the NA49 experiment at SPS in this figure, for two reasons.  Firstly, the balance function in that experiment was not measured at mid-rapidity.  Secondly, the non-uniform acceptance in pseudorapidity makes the simplified correction of Eq.~\ref{Eq:AcceptanceCorrection} invalid.}

\begin{figure}[thb!]
\centering
\includegraphics[width=0.92\linewidth]{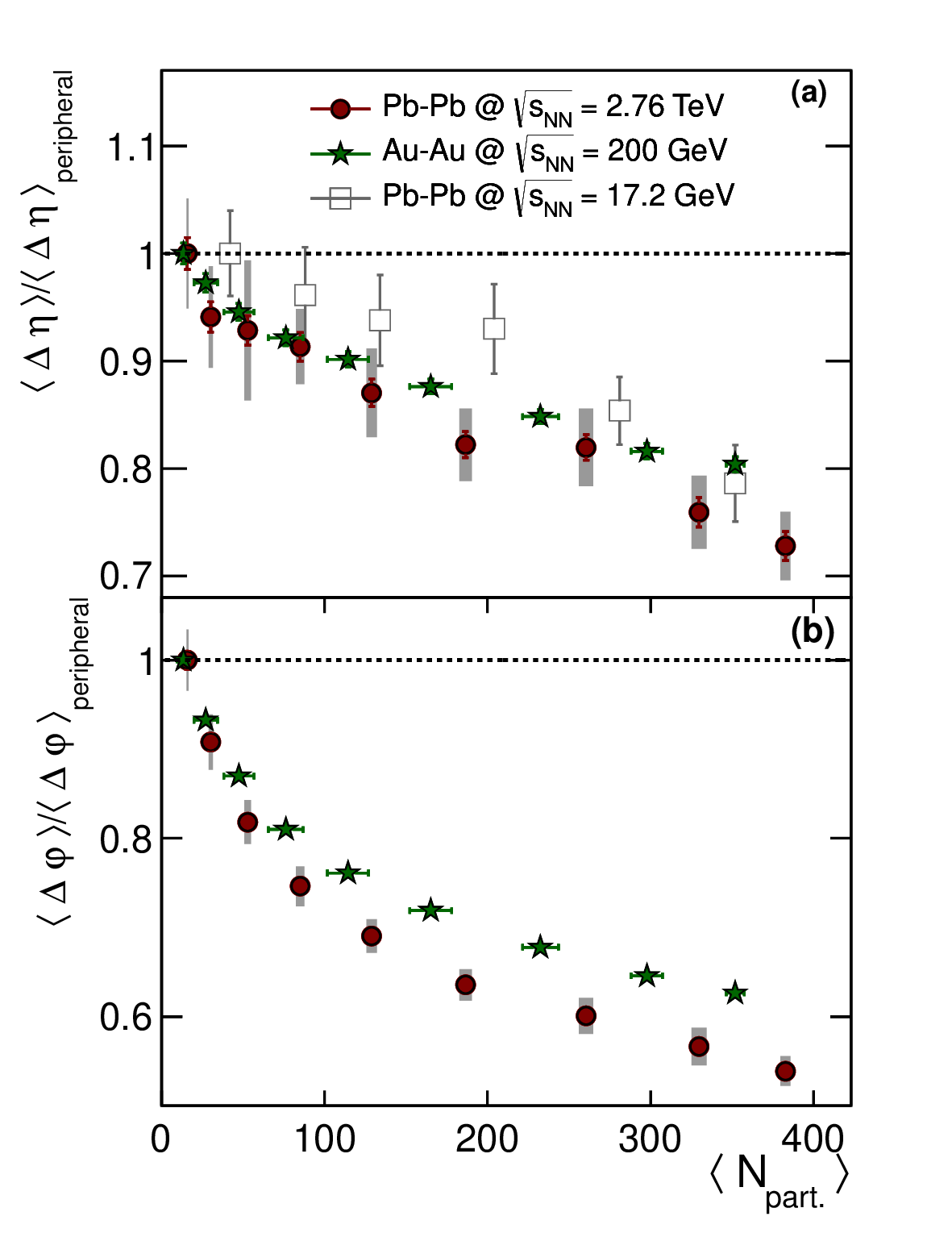}
\caption{(Color online).  The centrality dependence of the relative decrease of the width of the balance 
function in the relative pseudorapidity (a) and relative azimuthal angle (b). The ALICE points are compared to results for the highest SPS \cite{Ref:BalanceFunctionNA49} and RHIC \cite{Ref:BalanceFunctionSTAR2} energies.}
\label{fig:StarComparisonCP}
\end{figure}

While the centrality dependence is similar for both measurements, the widths are seen to be significantly narrower at the LHC energies. This is consistent with the idea of having a system exhibiting 
larger radial flow at the LHC with respect to RHIC \cite{Ref:AliceFlow}  while having a longer-lived QGP 
phase \cite{Ref:AliceHBT} with the consequence of a smaller separation between charge pairs when created at hadronization. However, it is seen that the relative decrease of the width between central and peripheral collisions seems to be similar between the two energies. This observation could challenge the interpretation of the narrowing of the width in \deta~as primarily due to the late stage creation of balancing charges.

To further quantify the previous observation, Fig.~\ref{fig:StarComparisonCP} presents the relative decrease of $\langle \Delta \eta \rangle$ 
(a) and $\langle \Delta \varphi \rangle$ (b) from peripheral to central collisions as a function of the mean number of participating nucleons, $\langle N_{\mathrm{part}} \rangle$, for the highest SPS\footnote{We include the NA49 points in this representation since the ratio to the peripheral results should cancel out the acceptance effects to first order.} \cite{Ref:BalanceFunctionNA49} and RHIC \cite{Ref:BalanceFunctionSTAR2} energies, compared to the values reported in this Letter. In this figure, central (peripheral) collisions correspond to high (low) number of $\langle N_{\mathrm{part}} \rangle$. The choice of the representation as a function of $\langle N_{\mathrm{part}} \rangle$ is mainly driven by the apparent better scaling compared to the centrality percentile. It is seen that in terms of correlations in relative 
pseudorapidity the data points at the different energies are in fairly good agreement within the uncertainties, resulting though into an 
additional, marginal decrease for the 0-5$\%$ most central collisions of  $\approx (9.5 \pm 2.0\,(stat) \pm 2.5\,(syst))\%$ compared to the RHIC point.
On the other hand, $\langle \Delta \varphi \rangle/\langle \Delta \varphi \rangle_{peripheral}$ exhibits a decrease of $\approx (14.0 \pm 1.3\,(stat) \pm 1.9\,(syst))\%$ between the 
most central Au--Au collisions at $\sqrt{s_{\rm NN}} = 200$~GeV and the results reported in this Letter. 
This could be attributed to the additional increase in radial flow between central and peripheral 
collisions at the LHC compared to RHIC energies. Another contribution might come from the 
bigger influence from jet-like structures at the LHC with respect to RHIC that results in particles 
being emitted preferentially in cones with small opening angles.
Contrary to $\langle \Delta \varphi \rangle/\langle \Delta \varphi \rangle_{peripheral}$, this strikingly marginal decrease of $\langle \Delta \eta \rangle/\langle \Delta \eta \rangle_{peripheral}$ between the three colliding energy regimes that differ more than an order of magnitude, cannot be easily understood solely within the framework of the late stage creation of charges. 

\section{Summary}
\label{Section:Summary}

This Letter reported the first measurements of the balance function for charged particles in Pb--Pb 
collisions at the LHC using the ALICE detector. The balance function was studied both, in relative pseudorapidity 
($\Delta \eta$) and azimuthal angle ($\Delta \varphi$). The widths of the balance functions, \wdeta~and $\langle \Delta \varphi \rangle$, are found to decrease 
when moving from peripheral to central collisions. The results are consistent with the picture of a system 
exhibiting larger radial flow in central collisions but also whose charges are created at a later stage of the collision. While HIJING is not able to reproduce the observed centrality dependence of the width in either projection, 
AMPT tuned to describe the v$_2$ values reported by ALICE seems to agree qualitatively with the centrality 
dependence of \wdphi~but fails to reproduce the dependence of $\langle \Delta \eta \rangle$. 
A thermal blast-wave model incorporating the principle of local charge conservation was fitted to the centrality 
dependence of \wdeta~and $\langle \Delta \varphi \rangle$. The resulting values of the charge separation at freeze-out can be used to constrain models describing the hadronization processes. 
The comparison of the results with those from lower energies showed that the centrality dependence of the width, in both the relative pseudorapidity and azimuthal angle, when scaled by the most peripheral widths, exhibits minor differences between RHIC and LHC.

These studies will soon be complemented by and extended to the correlations of identified particles in an 
attempt to probe the chemical evolution of the produced system, to quantify the influence of radial flow to the narrowing of the balance function width in more central collisions and to further constrain the parameters 
of the models used to describe heavy-ion collisions.

\ifpreprint
\iffull
\newenvironment{acknowledgement}{\relax}{\relax}
\begin{acknowledgement}
\section*{Acknowledgements}
We would like to thank Scott Pratt for providing us with the blast--wave model calculations and fruitful discussions.
The ALICE Collaboration would like to thank all its engineers and technicians for their invaluable contributions to the construction of the experiment and the CERN accelerator teams for the outstanding performance of the LHC complex.
\\
The ALICE Collaboration acknowledges the following funding agencies for their support in building and
running the ALICE detector:
 \\
State Committee of Science, Calouste Gulbenkian Foundation from
Lisbon and Swiss Fonds Kidagan, Armenia;
 \\
Conselho Nacional de Desenvolvimento Cient\'{\i}fico e Tecnol\'{o}gico (CNPq), Financiadora de Estudos e Projetos (FINEP),
Funda\c{c}\~{a}o de Amparo \`{a} Pesquisa do Estado de S\~{a}o Paulo (FAPESP);
 \\
National Natural Science Foundation of China (NSFC), the Chinese Ministry of Education (CMOE)
and the Ministry of Science and Technology of China (MSTC);
 \\
Ministry of Education and Youth of the Czech Republic;
 \\
Danish Natural Science Research Council, the Carlsberg Foundation and the Danish National Research Foundation;
 \\
The European Research Council under the European Community's Seventh Framework Programme;
 \\
Helsinki Institute of Physics and the Academy of Finland;
 \\
French CNRS-IN2P3, the `Region Pays de Loire', `Region Alsace', `Region Auvergne' and CEA, France;
 \\
German BMBF and the Helmholtz Association;
\\
General Secretariat for Research and Technology, Ministry of
Development, Greece;
\\
Hungarian OTKA and National Office for Research and Technology (NKTH);
 \\
Department of Atomic Energy and Department of Science and Technology of the Government of India;
 \\
Istituto Nazionale di Fisica Nucleare (INFN) and Centro Fermi -
Museo Storico della Fisica e Centro Studi e Ricerche "Enrico
Fermi", Italy;
 \\
MEXT Grant-in-Aid for Specially Promoted Research, Ja\-pan;
 \\
Joint Institute for Nuclear Research, Dubna;
 \\
National Research Foundation of Korea (NRF);
 \\
CONACYT, DGAPA, Mexico, ALFA-EC and the HELEN Program (High-Energy physics Latin-American--European Network);
 \\
Stichting voor Fundamenteel Onderzoek der Materie (FOM) and the Nederlandse Organisatie voor Wetenschappelijk Onderzoek (NWO), Netherlands;
 \\
Research Council of Norway (NFR);
 \\
Polish Ministry of Science and Higher Education;
 \\
National Authority for Scientific Research - NASR (Autoritatea Na\c{t}ional\u{a} pentru Cercetare \c{S}tiin\c{t}ific\u{a} - ANCS);
 \\
Ministry of Education and Science of Russian Federation,
International Science and Technology Center, Russian Academy of
Sciences, Russian Federal Agency of Atomic Energy, Russian Federal
Agency for Science and Innovations and CERN-INTAS;
 \\
Ministry of Education of Slovakia;
 \\
Department of Science and Technology, South Africa;
 \\
CIEMAT, EELA, Ministerio de Educaci\'{o}n y Ciencia of Spain, Xunta de Galicia (Conseller\'{\i}a de Educaci\'{o}n),
CEA\-DEN, Cubaenerg\'{\i}a, Cuba, and IAEA (International Atomic Energy Agency);
 \\
Swedish Research Council (VR) and Knut $\&$ Alice Wallenberg
Foundation (KAW);
 \\
Ukraine Ministry of Education and Science;
 \\
United Kingdom Science and Technology Facilities Council (STFC);
 \\
The United States Department of Energy, the United States National
Science Foundation, the State of Texas, and the State of Ohio.
\end{acknowledgement}
\ifbibtex
\bibliographystyle{utphys}
\bibliography{biblio}{}
\else

\fi
\newpage
\appendix
\section{The ALICE Collaboration}
\label{app:collab}
\else
\ifbibtex
\bibliographystyle{utphys}
\bibliography{biblio}{}
\else

\fi
\fi
\else
\iffull
\vspace{0.5cm}

\else
\ifbibtex
\bibliographystyle{utphys}
\bibliography{biblio}{}
\else

\fi
\fi
\fi
\end{document}